\documentclass[aps,twocolumn,floatfix,superscriptaddress]{revtex4}
\usepackage[dvips]{graphicx,rotating,epsfig}
\usepackage{indentfirst}
\usepackage{amsmath,amssymb}
\usepackage{times}

\begin{document}

\title{Molecular Dynamics Simulations of Detonation Instability}

\author{Andrew J. Heim}

\affiliation{Department of Applied Science,
University of California, Davis, California 95616}

\affiliation{Theoretical Division,
Los Alamos National Laboratory, Los Alamos, New Mexico, 87545}

\author{Niels Gr{\o}nbech-Jensen}

\affiliation{Department of Applied Science,
University of California, Davis, California 95616}

\author{Edward M. Kober}

\author{Timothy C. Germann}

\affiliation{Theoretical Division,
Los Alamos National Laboratory, Los Alamos, New Mexico, 87545}

\date{\today}

\begin{abstract}
After making modifications to the Reactive Empirical Bond Order potential for 
Molecular Dynamics (MD) of Brenner \emph{et al.} in order to make the model 
behave in a more conventional manner, we discover that the new model exhibits 
detonation instability, a first for MD\@. The instability is analyzed in 
terms of the accepted theory. 
\end{abstract}

\maketitle
%Introduction

\section{Introduction}
There are obvious problems with the 1D theory of detonation. Foremost is that 
it is impossible to create truly 1D detonation experiments. In many cases 
experiments come close, and their results can be extrapolated to match theory. 
But for certain classes of explosives, namely gas phase ones, the shape of the 
shock front, no matter how one-dimensional the experimental setup is, is 
intrinsically 3D with ripples. These detonations leave patterns in the soot 
on the walls of shock tubes, indicating the incidence of a section of 
the front with high vorticity with the wall. Such detonations are said to be 
unstable. See for example Chapter~7 of Fickett and Davis's book 
\cite{Fickett} or Austin's dissertation \cite{Austin}. 

The first truly successful work in theoretically analyzing 
this phenomenon was done by Erpenbeck in a series of papers published 
throughout the sixties~\cite{Erpenbeck1,ErpenbeckII,ErpenbeckIII,ErpenbeckIV} 
(summarized in chapter~6 of ref.~\cite{Fickett}) in which he 
describes a Laplace transform procedure for determining stability for a given 
parameterization of the equation of state (EOS) and reaction rate of a 
material. In 1964 he applied 
his formulation to 1D instability for a variety of parameterizations 
\cite{ErpenbeckII}, and in 1966 he expanded to transverse perturbations 
(multi-D) \cite{ErpenbeckIII}. 
In 1970 he summarized his decade of work \cite{ErpenbeckIV}.

Although Erpenbeck's method was the 
only one at the time to definitively show whether a parameterization was 
stable, it was not widely used for reasons mentioned by Lee and Stewart 
\cite{Lee}. 
The stability of a detonation based on a given 
parameterization had to be figured point-wise and the stability boundary 
interpolated. Lee and Stewart introduced a normal mode analysis via 
shooting method \cite{Lee} that is easier and faster than Erpenbeck's Laplace 
transform method. The normal mode analysis has fewer drawbacks about in which 
regimes it is useful, namely underdriven detonations. 

After the introduction of normal mode analysis, there 
was an expansion in detonation instability research. Notable is the work of 
Bourlioux and Majda \cite{Bourlioux,Majda}, who were the first to compare 
the results of the normal mode analysis with hydrodynamic detonation 
simulations in 2D\@. 
Short and Stewart extended the work of Lee and Stewart to transverse 
perturbations \cite{Stewart}. 

Most analysis and simulation have been on 
square detonations (instantaneous shock rise, finite induction zone, and 
instantaneous heat release, typical of detonations of high activation 
energy \cite{Fickett}) 
with a polytropic gas EOS and an Arrhenius reaction rate with 
one-step reactions (although other rates, numbers of reaction steps, and 
different EOSs have been examined \cite{Short}). 
The main parameters of concern in these analyses are the 
activation energy, exothermicity, adiabatic gamma, and degree of overdrive. 
Short gives a good summary of the research done since normal mode analysis in 
the area of detonation instability \cite{Short}. 
Two notable observations are that all of the 
simulations are done by continuum hydrodynamics codes and that condensed phase 
high explosives are, for the most part, stable. Instability has never been 
shown in the current range of limited molecular dynamics (MD) studies, which 
make no assumptions about the EOS and reaction rate. 

In the preceding paper \cite{Heim2} we justify modifications to the 
functional form of 
a popular MD model (ModelI) for high explosives (HEs) (see Eq.~1 and Table I in 
the errata of Brenner \emph{et al}.~\cite{Brenner}). The result is a new 
potential (ModelIV), which behaves in a more conventional manner. We notice, 
however, that the measured shock velocity is noticeably faster than the 
Chapman--Jouguet (CJ) value for an underdriven detonation. Where the 
preceding paper \cite{Heim2} focuses mainly on the difference in 
thermo-chemical properties of the 
two models, this current work is concerned with the results of non-equilibrium 
MD (NEMD) simulations of the ModelIV potential, which will demystify the shock 
velocity discrepancy. 
In Sec.~\ref{sec:param} we explain our choice of dissociation energies. 
In Sec.~\ref{sec:instability} the detonation instability is measured and 
analyzed in terms of theory. 

%Section Parameterization
\section{Parameterization}
\label{sec:param}
 
Two types of 2D simulations using the ModelIV model are performed. 
The preceding paper \cite{Heim2} discusses 
the results of microcanonical simulations for seeking the CJ state and 
conducting Arrhenius cookoffs. There a parameterization of 
ModelIV in which the metastable dissociation energy $D_e^{\textrm{AB}}=1.0$~eV 
and the stable dissociation energy $D_e^{\textrm{AB}}=5.0$~eV such that the 
exothermicity $Q=D_e^{\textrm{AA/BB}}-D_e^{\textrm{AB}}=4.0$~eV is used, 
whereas with ModelI both $D_e^{\textrm{AB}}$ and $Q$ were never varied away 
from the REBO defaults \cite{Brenner} at once. 
This parameterization was initially chosen to create a self-propagating 
detonation front. In this section the results of NEMD simulations are 
discussed, in which supported and overdriven detonations are modeled. 
The reason for the choice of parameterization is discussed. 

Initially, the same bonding energies as with 
ModelI, $D_e^{\textrm{AB}}=2.0$~eV and $D_e^{\textrm{AA/BB}}=5.0$~eV, are used. 
The first characteristic discovered is that the new material is insensitive to 
shock initiation. In fact, initiation requires a rigid piston driving at a 
velocity of about 4.5 km/s, an overdriven state. Since it is presumed 
that this new material has a high activation energy, but sufficient 
exothermicity to 
maintain detonation once initiated even if unsupported, the piston's velocity 
is backed 
off to zero. The result is that the detonation quickly quenches once the 
rarefaction wave that emanates from the decelerated piston catches up to the 
reaction zone. It seems, then, that we do not have a true HE\@. 

It should be 
noted that even with a strong overdriving piston, the material at the front 
demonstrates a resilience not found in ModelI\@. Where ModelI succumbs to 
nearly 
instantaneous dissociation upon compression by an unsupported detonation, 
ModelIV 
demonstrates uniaxial compression in a finite induction zone, 
characterized by vertically aligned (detonation travels in the horizontal 
direction) reactant dimers before dissociation.

\begin{figure}
\includegraphics*[width=\linewidth]{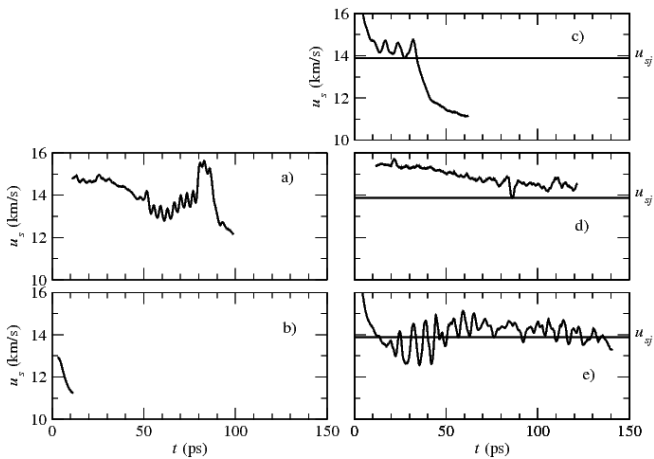}
\caption[]{Shock velocity vs.~time. In the first column $Q=3.0$~eV and  
4.0~eV in the second. Constant lines in the second column indicate the CJ 
value of $u_s$. This is not known for $Q=3.0$~eV\@. Rows indicate the 
distance between periodic boundaries in the transverse direction ($w_x$). 
For row one $w_x = 24~l_x$, where $l_x$ is the unit lattice parameter. For 
row two $w_x = 96~l_x$, and $w_x = 192~l_x$ in row three. 
Simulations represented by panes a through c are known to fail, although 
simulation b may not have had sufficient initiation.  
\label{fig:usvt}}
\end{figure}

\begin{figure*}
  %\setlength{\unitlength}{\textheight}
  %\begin{picture}(.5,.85)(0,0)
  %  \put(.02,.02){
      \includegraphics*[width=\textwidth]%height=.8\unitlength]
		       {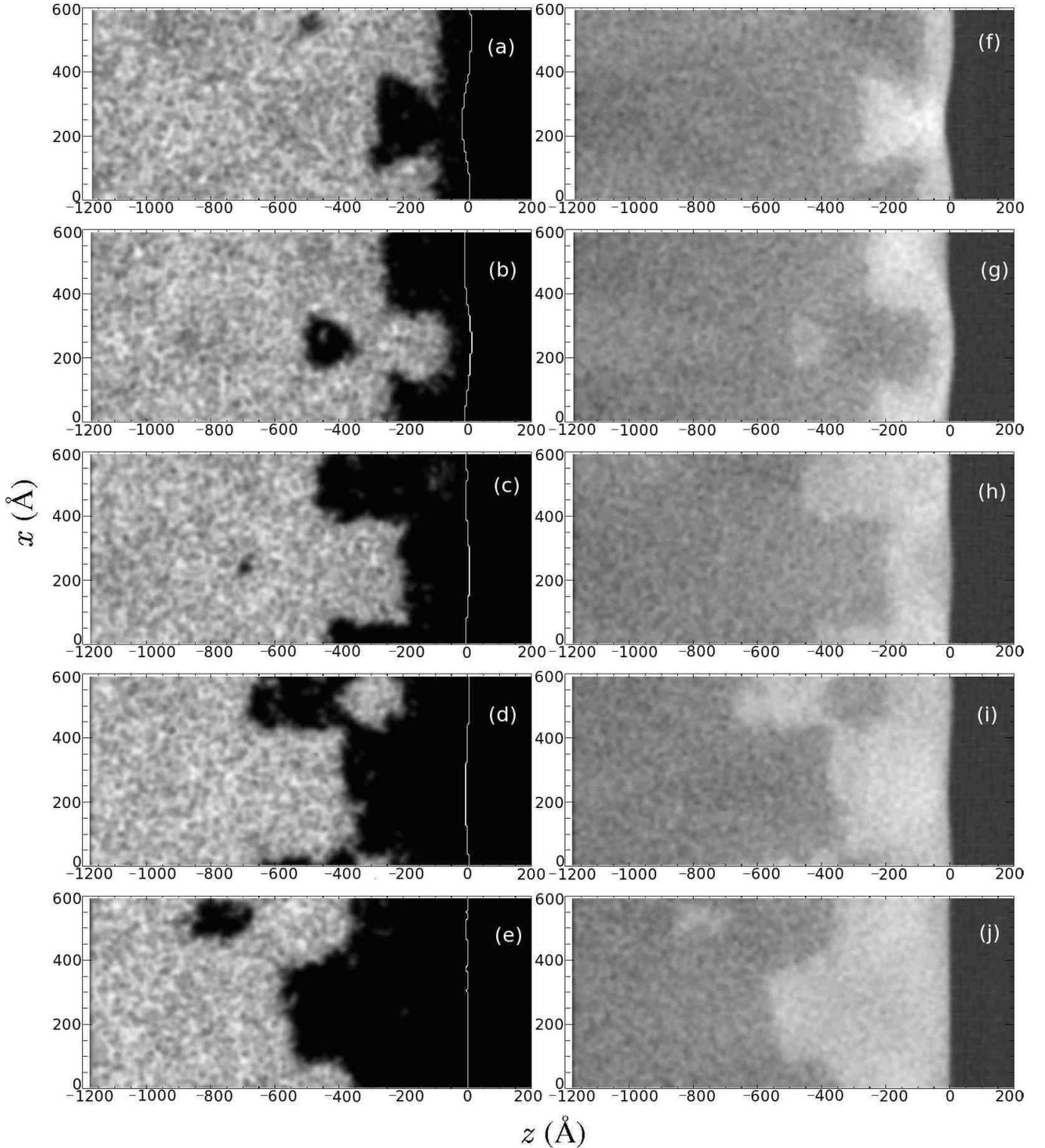}
    %}
    %\put(0,.4){
    %  \begin{rotate}{90}
%	$x$~(\AA)
%      \end{rotate}
%    }
%    \put(.25,0){$z$~(\AA)}
%  \end{picture}
\caption[Time evolution of the degree of reaction (left column), 
defined as the fraction of particles in the product state, and number density 
(right column) for the 
parameterization in which $D_e^{\textrm{AB}}=1.0$~eV and 
$D_e^{\textrm{AA/BB}}=4.0$~eV\@.]
{Time evolution of the degree of reaction (left column), 
defined as the fraction of particles in the product state, and number density 
(right column) for the parameterization in which $D_e^{\textrm{AB}}=1.0$~eV 
and $D_e^{\textrm{AA/BB}}=4.0$~eV\@.  
White lines indicate the position of the shock front, as determined for 
each value of $x$ by the steepest gradient of the $z$ component of particle 
velocity. Values increase 
from black to white. From top down, each row is at time $t=$ 82.7, 
85.6, 88.4, 91.3, and 94.1 ps. 
As the detonation fails, the reaction zone pulls away 
from the shock front, which flattens.
\label{fig:eta_n_lam_sample_ne3}}
\end{figure*}

\begin{figure}
  %\setlength{\unitlength}{0.9\linewidth}
  %\begin{picture}(1,.7)(0,0)
  %  \put(.00,.0){
      \includegraphics*[width=\linewidth]%height=.7\unitlength]
		       {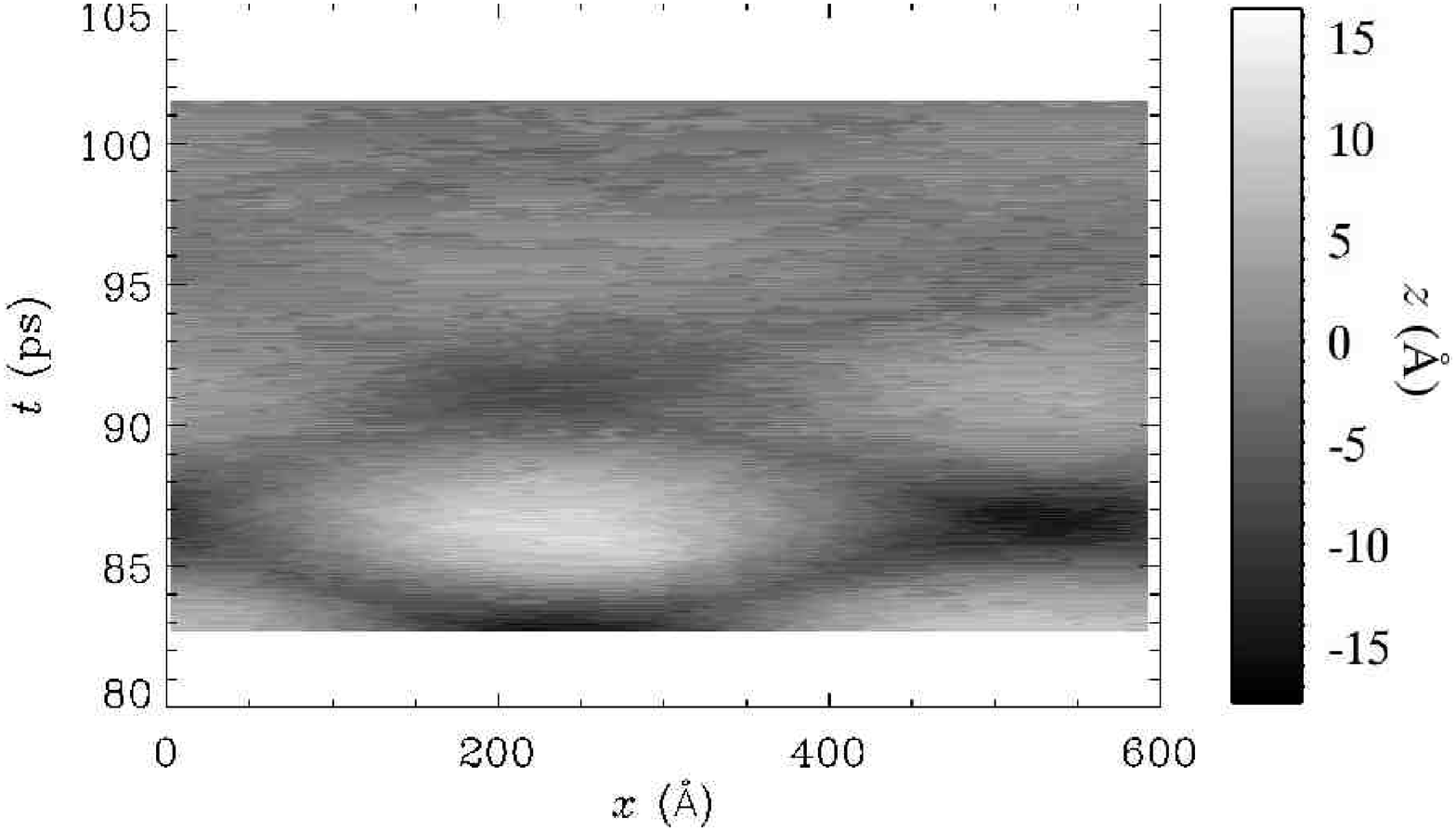}
   % }
   % \put(0,.32){
   %   \begin{rotate}{90}
	%$t$~(ps)
%      \end{rotate}
%    }
    %\put(.45,0){$x$~(\AA)}
%  \end{picture}
\caption[Contour plot of the shock front for the parameterization in which 
$D_e^{\textrm{AB}}=1.0$~eV and $D_e^{\textrm{AA/BB}}=4.0$~eV\@.]
{Contour plot of the local position of the shock front relative to 
the average over the transverse direction for the parameterization in which 
$D_e^{\textrm{AB}}=1.0$~eV and $D_e^{\textrm{AA/BB}}=4.0$~eV\@.
Black is behind the average and 
white is in front. Notice that the amplitude decreases with time.
\label{fig:frnt_shp_cntr_ne3}}
\end{figure}

Since it is assumed that the exothermicity is sufficient, the parameters for 
a second simulation are next set to 
\linebreak[1] $D_e^{\textrm{AB}}=1.0$~eV and $D_e^{\textrm{AA/BB}}=4.0$~eV 
such that $Q$ still is 3.0~eV\@. This 
detonation, which, as with the previous parameterization, is initiated by an 
overdriving piston that was backed off to zero, eventually fails too (see pane 
a in Fig.~\ref{fig:usvt}).
Upon inspection of Fig.~\ref{fig:eta_n_lam_sample_ne3}, 
one will notice that the front seems to have a standing wave shape, but 
Fig.~\ref{fig:frnt_shp_cntr_ne3} shows that there are 
symmetric traveling waves, 
probably emanating from either side of the same event. 

Structure in the front is a 
tell-tale sign of instability in the detonation and violates the 1D assumption 
of the Zel'dovich, von Neumann, and D{\"o}ring theory \cite{Fickett}. 
Such structure has been theorized 
\cite{Erpenbeck1} and shown through experiments and hydrodynamic 
simulations---mostly in reactive gases---\cite{Fickett} to be caused by 
transverse waves 
crossing in the reaction zone. The keystone shape in the $\lambda$ contour in 
the first row of Fig.~\ref{fig:eta_n_lam_sample_ne3} is 
also typical of such unstable detonations in gases \cite{Austin}. 

The number density ($\eta$) contours in Fig.~\ref{fig:eta_n_lam_sample_ne3} 
are plotted in order to see these postulated transverse waves, but the results 
are inconclusive. 
Notice that the degree of reaction ($\lambda$) and $\eta$ seem to have 
an inverse relationship within the reaction zone. Behind a trough in the 
shock front, $\eta$ is high and $\lambda$ is low. Eventually the material 
there reacts. The resulting energy release pushes the products out of the area 
thus surging the trough forward. 

Since it is hypothesized that the distance between the periodic boundaries in 
the transverse direction dictates the wavelength of what seemed to be a 
standing wave,  
a third simulation with all of the same parameters, but with a sample 
that is twice as wide (192 lattice cells as opposed to 96), is started. 
This simulation quenches quickly (see pane b in Fig.~\ref{fig:usvt}) 
so a fourth is performed with 
an initiating piston that starts out moving fifty percent faster, but 
that, too, does not sustain the detonation. 
If there are transverse waves, they may help to sustain the 
detonation in the thinner simulation. Increasing the distance between the 
boundaries may have dropped the frequency of crossings of the transverse waves 
below a critical limit. This also implies that the main shock front is 
insufficient to sustain detonation. 

\begin{figure}
%  \setlength{\unitlength}{0.9\linewidth}
%  \begin{picture}(1,.7)(0,0)
%    \put(.02,.03){
      \includegraphics*[width=\linewidth]%height=.68\unitlength]
		       {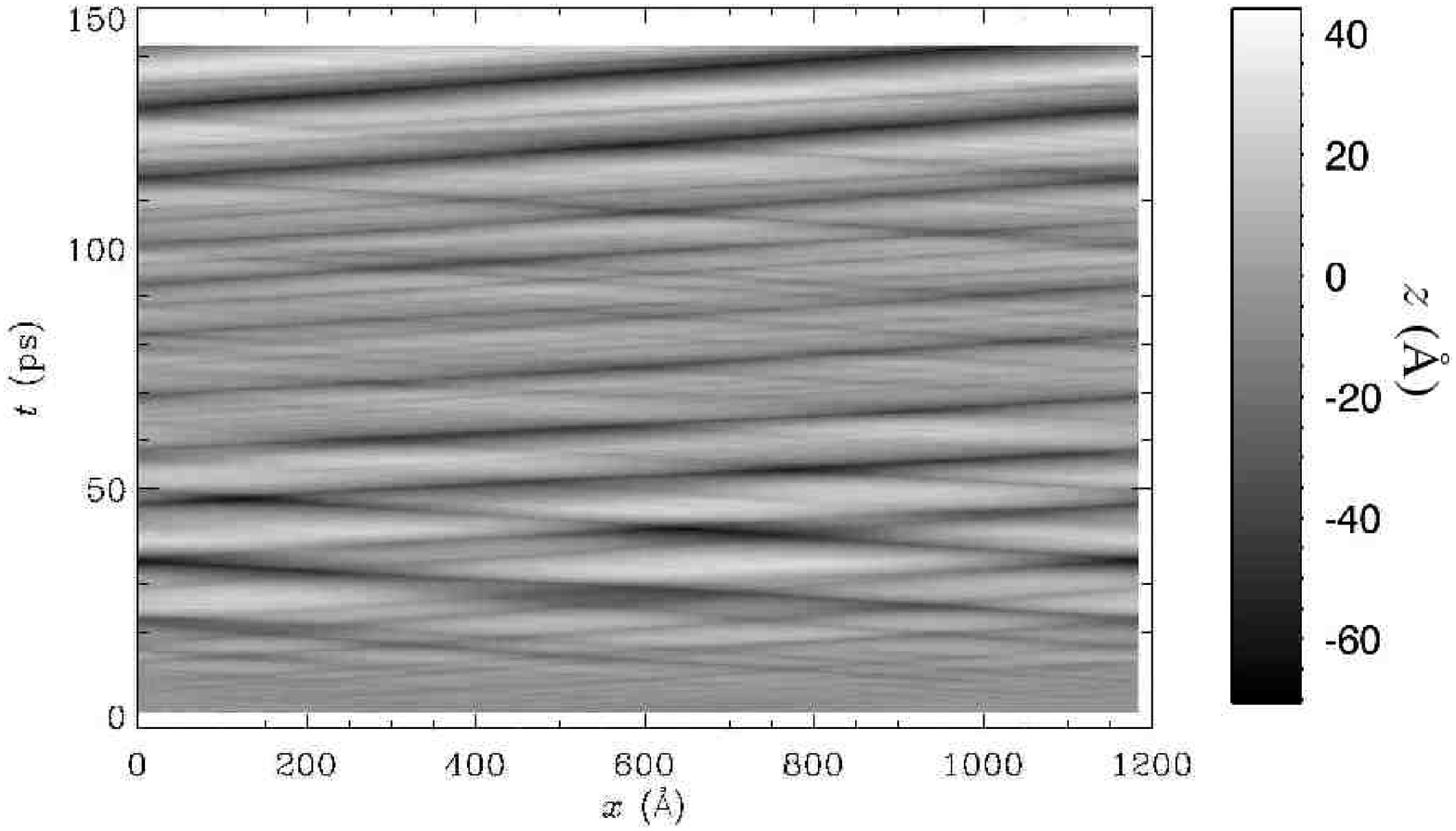}
%    }
%    \put(0,.32){
%      \begin{rotate}{90}
%	$t$~(ps)
%      \end{rotate}
%    }
%    \put(.45,0){$x$~(\AA)}
%  \end{picture}
\caption[Contour plot of the local position of the shock front relative to 
the average over the transverse direction for the parameterization in which
$D_e^{\textrm{AB}}=1.0$~eV and $D_e^{\textrm{AA/BB}}=5.0$~eV\@.]
{Contour plot of the local position of the shock front relative to 
the average over the transverse direction for the parameterization in which
$D_e^{\textrm{AB}}=1.0$~eV and $D_e^{\textrm{AA/BB}}=5.0$~eV\@. 
Black is behind the average and white is in front.
\label{fig:frnt_shp_cntr_ne4}}
\end{figure}

In order to make the detonation be self sustaining, a third parameterization 
in which $D_e^{\textrm{AB}}=1.0$~eV and 
$D_e^{\textrm{AA/BB}}=5.0$~eV, such that $Q=4.0$~eV, is tried. So far this 
simulation, which, as with the preceding, is initiated with an overdriving 
piston that backs off to zero, has not shown signs of failure (see pane e in 
Fig.~\ref{fig:usvt}). It is with 
this parameterization that this research proceeds. This simulation 
is as thick as the thicker of the two previous (192 lattice cells thick). 
In this simulation the front has an asymmetric structure 
(see Fig.~\ref{fig:frnt_shp_cntr_ne4})\@. 
%

%Detonation Instability
\section{Detonation Instability}
\label{sec:instability}

\begin{figure*}
  %\setlength{\unitlength}{\textheight}
  %\begin{picture}(.5,.85)(0,0)
  %  \put(.02,.02){
      \includegraphics*[width=\textwidth]%height=.7\textheight]
	{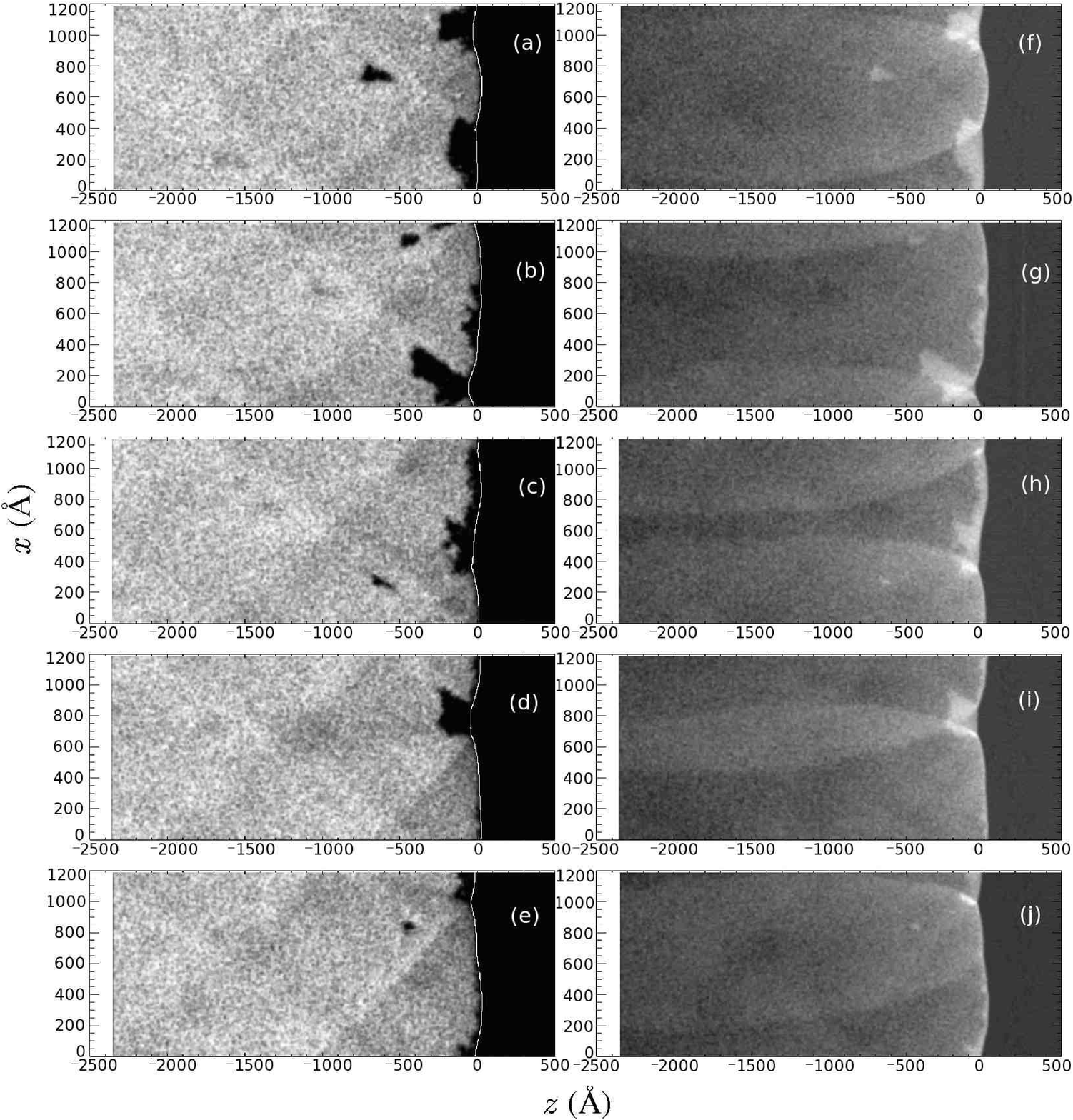}
      %lam_selection_ne4.ps}
    %}
    %\put(0,.4){
    %  \begin{rotate}{90}
	%$x$~(\AA)
    %  \end{rotate}
    %}
    %\put(.25,0){$z$~(\AA)}
  %\end{picture}
\caption[Time evolution of the degree of reaction (left column),  
and the number density (right column)\@ for the 
parameterization in which $D_e^{\textrm{AB}}=1.0$~eV and 
$D_e^{\textrm{AA/BB}}=5.0$~eV\@.]
{Time evolution of the degree of reaction (left column), defined as the 
fraction of particles in the product state,  
and the number density (right column) for the 
parameterization in which $D_e^{\textrm{AB}}=1.0$~eV and 
$D_e^{\textrm{AA/BB}}=5.0$~eV\@.  
White lines indicate the position of the shock front. Values increase 
from black to white. From top down, each row is at time $t=$ 44.5, 47.4, 
50.3, 53.1, and 56.0 ps.  
\label{fig:eta_n_lam_sample_ne4}}
\end{figure*}

Notice the distinct cellular structure in the $\lambda$ contours, especially 
in the latter frames of Fig.~\ref{fig:eta_n_lam_sample_ne4}. 
This is typical of unstable detonation and is 
seen in soot prints left on the walls of rectangular tubes used in detonation 
experiments and seen in certain properties in hydrodynamic 
simulations of the reactive Euler equations \cite{Fickett}\@. 
Where experiments using condensed phase explosives show a 
break down of this regular cell structure, ModelIV has a clear cell structure. 
There are many differences between ModelIV and real condensed phase 
explosives, 
particularly in the current implementation, which is performed on perfect 
crystals while real materials have a high concentration of defect structures. 
Any of said differences could mean the difference between cellular 
structure and random structure, but ModelIV suggests there is more of a 
reason for the real break down than merely being in the condensed phase. 

From the $\eta$ contours (Fig.~\ref{fig:eta_n_lam_sample_ne4}), 
one can clearly see transverse waves (perhaps so here and not in 
Fig.~\ref{fig:eta_n_lam_sample_ne3} because here the resolution is finer and 
$Q$ is greater so that the gradients at the fronts of the transverse waves are 
also greater). The sequence of events displayed in 
Fig.~\ref{fig:eta_n_lam_sample_ne4} 
can be matched to the accepted 
theory, of which Fickett and Davis give a good summary in 
Chapter~7 of their book \cite{Fickett}. 

Following their example, start with a flat shock wave, 
followed by an induction zone 
and then a reaction zone, a 1D situation. If a point  
``microdetonation'' occurs within the induction zone, it will produce a shock 
wave of its own, the shape of which will be skewed by the rarefaction and flow 
already present. It will have an arched shape. Since the shock stability 
conditions 
state so, the forward facing part of the microdetonation's shock will overtake 
the original flat front causing it to bulge forward. 

If there are two microdetonations, 
the transverse facing parts of their shock fronts will eventually meet. 
That 
intersection will move forward toward the shock front. Once the angle between 
the colliding transverse waves increases sufficiently, a Mach stem is formed. 
If conditions are right, this occurs within unreacted material, and their 
collision causes reaction, which surges the Mach stem forward of the incident 
shock front and intensifies the transverse waves as they pass through each 
other. 

As the Mach stem ages, the induction zone 
behind it lengthens because of rarefaction. The sections of transverse waves 
near the shock front move as 
detonation fronts through 
the rest of the induction zone, consuming post-shocked yet unburnt material on 
either side of the Mach stem. The transverse waves are not sustainable without 
fuel. If they again collide behind an old Mach stem (or any place with a long 
induction zone), they will have plenty of fuel to cause the formation of 
another rapid burn and forward surge. If a transverse wave encounters a newly 
formed Mach stem, behind which there is very little unburnt fuel, it will 
weaken. 

This gives spacing criteria for surviving transverse waves: the 
distance between them must be great enough that sufficient growth of the 
induction zone can take place between passings and small enough that new 
waves do not have sufficient space in which to grow and consume the fuel 
between existing waves. Since the NEMD simulations use periodic boundary 
conditions, the wave spacing is at most the simulation thickness. Since it 
was shown that these unsupported detonations cannot sustain themselves 
without the transverse waves, their ability to propagate depends on the 
simulation width as is evident in Fig.~\ref{fig:usvt}.   

The intersection of a transverse wave with the shock front is called the 
triple point. It is the passing of a triple point that etches cell patterns 
into the soot in shock tubes. It is also the source of the cell patterns 
seen in our $\lambda$ contours. Experiments cannot see cell patterns in 
chemistry, at least not in OH fluorescence from detonating O$_2$ + H$_2$ 
\cite{Austin}. Since, in thriving transverse waves, the triple 
point is crossing unreacted material and causing it to detonate, the  
shock front behind the triple point surges forward. The track of a triple 
point is thus seen as a traveling trough in contour plots of the relative 
shock position. 

If a transverse wave encountered only constant state material, it 
would reach a steady state. But, since there are variations by collisions 
with other transverse waves, passing through its own wake via reflection 
(or, in the case of the NEMD simulation, periodic boundaries), Mach stem 
formation, and thermal 
fluctuations, etc.~there can be a chaotic and stochastic quality to the 
behavior and number of transverse waves, the amount of which should depend on 
the transverse waves' sensitivity to such variations as well as the dimensions 
of the simulation. 
This dependence is evident in the degree of regularity in cell patters and 
and the evolution of the shock surface (see Fig.~\ref{fig:frnt_shp_cntr_ne4}). 

The example above is what one sees in 
Fig.~\ref{fig:eta_n_lam_sample_ne4} (and in the supplemental videos 
\footnote{See EPAPS Document Nos. [number will be inserted by publisher] and 
[number will be inserted by publisher]
for videos of $\eta$ and $\lambda$, which 
correspond to Fig.~\ref{fig:eta_n_lam_sample_ne4}\@. For more information on 
EPAPS, see {\tt http://www.aip.org/pubservs/epaps.html}.}). 
In the first $\eta$ frame (f) 
one can clearly see two transverse waves receding after the formation of 
a Mach stem, which has surged forward. The triple points of the transverse 
waves are headed toward unreacted material. Notice in the analogous 
$\lambda$ contour (a) that the induction zone is short behind the Mach stem. 
A final observation for this frame is that there is a plume of reactant that 
escapes detonation. 
It burns up more slowly as it flows downstream of the shock front. This is 
typical of unstable detonations \cite{Austin}. 

\begin{figure}
 \includegraphics*[width=\linewidth]{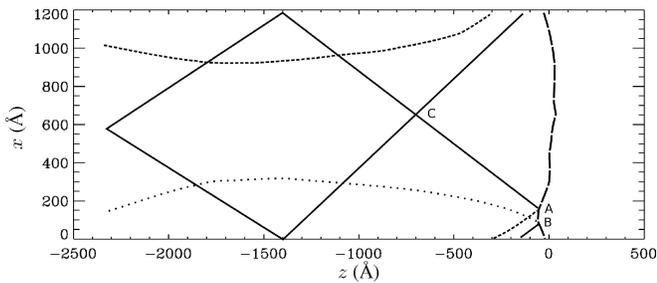}
 \caption[Trace of the relevant lines from the second row in 
Fig.~\ref{fig:eta_n_lam_sample_ne4} (frames b and g) at $t=$ 47.4 ps.]
{Trace of the relevant lines from the second row of 
Fig.~\ref{fig:eta_n_lam_sample_ne4} (frames b and g) at $t=$ 47.4 ps. 
Cell pattern and slip lines 
are solid lines. Shock front is a long-dashed line. Transverse 
wave for the triple point B is a dotted line. For A it is a 
short-dashed line. 
 \label{fig:eta_n_lam_lines}}
\end{figure}

In the second row of 
Fig.~\ref{fig:eta_n_lam_sample_ne4} (frames b and g), the 
triple points of the two transverse waves are about to collide 
in the lower half, 
where material is unreacted. The Mach stem is beginning to rarefy. We trace 
out the relevant lines from this row in Fig.~\ref{fig:eta_n_lam_lines}. 
In this Figure one can make out the two triple points, A and B, which are 
approaching each other. Between them is the incident front. On the line 
segments AC and BC, lie the slip lines for the respective triple points. 
Note that the BC slip line crosses the boundary. The slip lines mark the edge 
of the keystone shape. Here we have made it 
seem that the slip lines coincide with the cell pattern and thus the 
tracks of the triple points. The patterns are fuzzy and cannot be so 
accurately determined as to make such a claim. 

In row three of Fig.~\ref{fig:eta_n_lam_sample_ne4} (frames c and h), 
the material has combusted after the crossing of the triple points, 
and the front has surged forward, a state seen in the first row. 
A second Mach stem has formed while the original one 
has further rarefied, providing fuel for the yet again strengthened transverse 
waves. 

In rows four (d and i) and five (e and j) another Mach stem is formed, 
but the second one has not rarefied much yet. Because of a lack of fuel, 
the downward passing wave eventually disperses, as is 
evident in Fig.~\ref{fig:frnt_shp_cntr_ne4} by the disappearance of  
the dark line of negative slope shortly after 50~ps.

\section{Conclusion}

For ModelIV the activation energy is higher than with ModelI\@. The fact that 
instability appears in ModelIV but not ModelI follows the trend that stability 
tends to increase with decreasing activation energy \cite{Stewart}. 
As far as we know, this is the first time such instability has 
been demonstrated using an MD model. 

The hope is that the ability for MD to model instability will eventually 
reveal much about the 
phenomenon. It can be helpful because it needs no reaction rate or 
equation of state as input parameters and thus cuts down on some of the 
parameter space to be explored. For the former reason, it can be especially  
useful for 
the study of condensed phase high explosives since it is difficult not only to 
know their multiphase EOS throughout the reaction zone, but also to make real 
measurements in the reaction zone. Although this may not be too 
important for ZND like detonations, in which the unsupported shock velocity 
and the CJ state  
depend on the initial state and that of the products, it does become relevant 
in weak \cite{Fickett} detonations, in which the sonic point is at a 
state of incomplete reaction and which are a starting basis for instability 
analysis. 

If one thoroughly probes the EOS and the reaction rate 
of ModelIV, normal mode analysis could confirm its 
predictions.  The scaling of Short and Stewart \cite{Stewart} 
or one based on the CJ state may allow for comparison to real explosives. 

It should be noted that the cellular structure found with ModelIV is not 
typical of most condensed phase explosives in either experiment, continuum 
modeling, or normal mode analysis, yet this is not disheartening because of 
the aforementioned problems each of these techniques has with such explosives. 
Experiments may not be able to resolve them. Hydrodynamic codes and normal 
mode analysis may be using the wrong EOS or reaction rate, or ModelIV may be 
missing some characteristic of condensed phase explosives that tends to 
stabilize the detonations thereof \cite{Short}\@.

Many of these cited studies go into much detail about the 
parametric regimes in which their explosives are stable or not. They vary 
the activation energy, $Q$, the amount of overdrive of 
a piston, or the frequency or wavenumber of a perturbation for instance. Such 
a study would be worthwhile using ModelIV, but it is beyond the scope of the 
current research. However, this work has raised this as a significant issue, 
and shown that it can be studied with this approach. 

Of greatest significance 
is that this work has shown condensed phase systems with realistic molecular 
properties can sustain detonations with 2D instabilities. The net result is 
that the systems propagate with velocities somewhat higher than CJ and that 
their reaction zone structure is convoluted with the structure of the 
transverse waves. All analyses of condensed phase HEs presume that they are 
2D stable, as this is the only manner in which a complete analysis can be 
achieved. An example is the Cheetah EOS \cite{Fried} which presumes that the 
observed 
detonation velocity is in fact the true CJ velocity. The current work raises 
the possibility that these values may be several percent too high. Other 
features such as corner turning depend on the width of the effective reaction 
zone, where one must now consider how transverse waves may effect this. 
Consequently, it is important to have developed a tool with which to 
quantitatively explore these features.

\acknowledgments
The authors would like to thank Charles Kiyanda, Mark Short, Jerry Erpenbeck, 
Sam Shaw, Tariq Aslam, 
John Bdzil, Alejandro Strachan, Brad Holian, Tommy Sewell, Yogesh Joglekar, 
and David Hall for useful conversations. This material was 
prepared by the University of California under Contract W-7405-ENG-36 and 
Los Alamos National Security under Contract DE-AC52-06NA25396 with the 
U.S. Department of Energy. The authors particularly wish to recognize funding 
provided through the ASC Physics and Engineering Modeling program.

\end{document}